\documentclass{PoS}
\pdfoutput=1

\usepackage[utf8]{inputenc}
\usepackage{tikz}
\usepackage{amsmath}
\usepackage{amsfonts}
\usepackage{amssymb}
\usepackage{graphicx}
\usepackage{bbm}
\usepackage{braket}
\usepackage{xspace}
\usepackage{enumitem}

\newcommand{\Nf}{{N_f}}

\renewcommand{\epsilon}{\varepsilon}

\newcommand{\SU}{\text{SU}}

\newcommand{\dd}{D}
\newcommand{\one}{\mathbbm{1}}

\newcommand{\DOT}{.}
\newcommand\qone{QCD$_1$\xspace}
\newcommand\cc{\text{c.c.}}
\newcommand\ev[1]{\langle{#1}\rangle}

\newcommand{\Nmc}{N_\text{MC}}

\DeclareMathOperator{\re}{Re}

\DeclareMathOperator{\diag}{diag}
\DeclareMathOperator{\sign}{sign}
\DeclareMathOperator\arsinh{arsinh}
\DeclareMathOperator\tr{tr}

\title{Sign problem and subsets in one-dimensional QCD%
\thanks{Supported by DFG}}

\ShortTitle{Sign problem and subsets in one-dimensional QCD}

\author{\speaker{Jacques Bloch}, Falk Bruckmann, and Tilo Wettig\\
       Institute for Theoretical Physics, University of Regensburg, Germany\\
       E-mail: \email{jacques.bloch@ur.de}, \email{falk.bruckmann@ur.de}, \email{tilo.wettig@ur.de}}

\abstract{
We present a subset method that solves the sign problem for QCD at nonzero quark chemical potential in 0+1 dimensions. The subsets of gauge configurations are constructed using the center symmetry of the SU(3) group. These subsets completely solve the sign problem for up to five flavors. For a larger number of flavors the sign problem slowly reappears, and we propose an extension of the subsets that also solves the sign problem for these cases. The subset method allows for numerical simulations of the model at nonzero chemical potential. We also present some preliminary results on subsets for QCD in two, three, and four dimensions.
}

\FullConference{31st International Symposium on Lattice Field Theory - LATTICE 2013\\
		July 29 - August 3, 2013\\
		Mainz, Germany}
		
\begin{document}

\section{Introduction}

At nonzero chemical potential the determinant of the Dirac operator becomes complex and numerical simulations of QCD are hampered by the sign problem. Although this problem is particularly serious in four dimensions, it already exists in 0+1 dimensions \cite{Bilic:1988rw}. \qone can thus be used as a toy model to study the sign problem \cite{Ravagli:2007rw}.

As the sign problem in \qone is mild, standard reweighting methods could be used in numerical simulations. However, it would be interesting to find a solution to the sign problem as this could help to improve on this problem in higher dimensions.

One category of solutions to the sign problem can be called \textit{subset} methods, where configurations of the ensemble are gathered such that the sum of their complex weights is real and positive. If such subsets can be formed, they can be sampled in Monte Carlo (MC) simulations to construct Markov chains of relevant subsets with importance sampling and used to measure observables.

Subsets were already introduced before to solve the sign problem for the $q=3$ Potts model \cite{Alford:2001ug} and for a random matrix model of QCD \cite{Bloch:2011jx,Bloch:2012ye}. 

The existence of such subsets is often related to some symmetry of the model, and in the case of the random matrix model the subsets are closely related to a projection on the $q=0$ canonical determinant \cite{Bloch:2012bh}. 
The same subset construction can be used to solve the sign problem in U($N_c$) theories, but it is \textit{not} allowed in QCD because the configurations in the subsets would leave SU(3). 

Therefore another subset construction is needed for QCD, and the idea presented in this talk is to construct subsets based on the $Z_3$ center symmetry of SU(3). More details about the subset construction and results can be found in Ref.~\cite{Bloch:2013ara}.

\section{Sign problem in QCD in 0+1 dimensions}

We consider \qone, which is an SU(3) gauge theory on one spatial point and $N_t=1/aT$ time slices. 
The \qone Dirac operator for a quark of mass $m$ at chemical potential $\mu$ is
\begin{align}
D_{tt'} =  
m \, \delta_{tt'} 
+ \frac{1}{2a}  
\left[ e^{a\mu} U_t \delta_{t',t+1} - e^{-a\mu} U_{t-1}^\dagger \delta_{t',t-1} \right]
 ,
\end{align}
where $U_t \in \SU(3)$ and $\delta_{tt'}$ is an anti-periodic Kronecker delta.
The Dirac determinant can be reduced to the determinant of a $3\times 3$ matrix \cite{Bilic:1988rw},
\begin{align}
  \det (aD)=\frac{1}{2^{3N_t}}\det \left[ e^{\mu/T} P+e^{-\mu/T} P^\dagger + 2\cosh\left(\mu_c/T\right)\,\one_3\right]
  \label{detD}
\end{align}
with Polyakov line $P= \prod_t U_t$ and effective mass $a\mu_c = \arsinh(am)$.
The determinant depends on $P$ and $\mu$ through the combination
$e^{\mu/T} P$ only because
(i) all gauge links can be shifted into the Polyakov line $P$ through an appropriate choice of gauge and
(ii) the $\mu$-dependence only arises from closed temporal loops. Hereafter we will set $a=1$.

As \qone has no gauge action the partition function $Z^{(N_f)}=\int dP\,{\det}^{N_f} D(P)$, with SU(3) Haar measure $dP$, is a one-link integral of the Dirac determinant for $N_f$ quark flavors.

Although the imaginary part of $\det^{N_f}D(P)$ can always be canceled by pairing $P$ with $P^*$ owing to $\det D(P^*)=[\det D(P)]^*$, the remaining real weight $\re\det^{N_f}D(P)$ has no fixed sign for $\mu \neq 0$ and a sign problem arises in MC simulations.

\section{Subset method}
\subsection{Subset construction}

The aim of the subset method is to gather configurations into small subsets such that the sum of their determinants is real and positive. For \qone we propose the following subset construction \cite{Bloch:2013ara}:
For any configuration $P$ a subset $\Omega_P \subset \SU(3)$ is constructed using $Z_3$ rotations and complex conjugation,
\begin{align}
  \Omega_P = \{ P\,, \,e^{2\pi i/3} P\,,\, e^{4\pi i/3} P \} \, \cup \, \{P \to P^*\} \,.
  \label{Z3subset}
\end{align}
Clearly, the set of all subsets forms a six-fold covering of the original SU(3) ensemble. 
If we define the subset weight as
\begin{align}
  \sigma(\Omega_P) = \frac16\sum_{k=0}^2 {\det}^{N_f} D(P_k) + \cc \,,\quad  P_k=e^{2\pi i k/3}P \,,
  \label{sigma}
\end{align}
the partition function can be rewritten as $Z^{(N_f)}=\int dP \,\sigma(\Omega_P)$.
If $\sigma(\Omega_P)$ is real and positive the partition function can be simulated using importance sampling.
As the subsets are generated with weights $\sigma(\Omega_P)$, ensemble averages of observables are computed as
\begin{align}
\langle O \rangle
= \frac{1}{Z^{(N_f)}} \int dP \,\sigma(\Omega_P) \, \ev{O}_{\Omega_P} 
\approx \frac1{\Nmc} \sum_{n=1}^{\Nmc} \ev{O}_{\Omega_{n}} ,
\end{align}
with subset measurements defined as
\begin{align}
  \langle O \rangle_{\Omega_P}= \frac{1}{6\sigma(\Omega_P)}
  \sum_{k=0}^{2} \left[ {\det}^{N_f} D(P_k) \, O(P_k) + (P_k \to P_k^*) \right]
  \label{subsetmeas}
\end{align}
to take into account that the configurations in a subset generically have different observable values.

\subsection{Subsets and zero triality}

The $N_f$-flavor determinant can be decomposed into powers of $e^{\mu/T}$ as
\begin{align}
{\det}^{N_f} D(P) = \sum_{q=-3N_f}^{3N_f} D_q \, e^{q\mu/T} .
\end{align}
Because the determinant \eqref{detD} satisfies
\begin{align}
 \det D(\underbrace{e^{i\theta}P}_\text{\hspace{-2cm}$Z_3$ rotation of $P$\hspace{-2cm}})\big|_{\mu/T}=\det D(P)\big|_{\underbrace{\scriptstyle\mu/T+i\theta}_{\scriptstyle\text{\hspace{-1cm}imaginary shift of $\mu$\hspace{-1cm}}}} 
\end{align}
the sum of the determinants in the $Z_3$ subsets corresponds to a projection on the zero triality sector,
\begin{align}
 \sigma(\Omega_P)=\frac{1}{3}\sum_{q=-3N_f}^{3N_f} \re\dd_q \, e^{q\mu/T} \underbrace{\sum_{k=0}^2
 e^{2\pi i qk/3}}_{\displaystyle 3\delta_{q\bmod 3,0}} 
 =\sum_{b=-N_f}^{N_f} \re\dd_{3b} \, e^{3b\mu/T}  \,,
\end{align}
which is now an expansion in the baryon number.

\section{Subset method for $\mathbf{\emph N_{\emph f}=1}$}

For $\Nf=1$ the subset weight is given by
\begin{align}
 \sigma(\Omega_P)= 2\cosh(3\mu/T) + A^3-3A+A|\tr P|^2 \quad \text{with} \quad A=2\cosh\left(\mu_c/T\right) \,.
\end{align}
The subset weight $\sigma(\Omega_P)$ is real and positive for any $\mu$, $m$, and $P$, and can be used to generate subsets with importance sampling.

We implemented the subset method and verified the results by comparing with known analytical predictions. The numerical algorithm consists of the following steps:
\begin{itemize}[topsep=0.3mm,itemsep=-1.5mm,leftmargin=3ex]

\item generate SU(3) links $P$ according to the Haar measure,

\item construct the $Z_3$ subsets $\Omega_P$ and explicitly compute the determinants and the subset weights,

\item perform a Metropolis accept-reject on the positive subset weights to construct a Markov chain of relevant subsets, 

\item compute the chiral condensate $\Sigma  = \frac{1}{N_t}\ev{\tr \left[D^{-1}\right]}$, quark number density $n = \frac{1}{N_t}\ev{\tr\left[D^{-1} \partial D/\partial\mu\right]}$, and average Polyakov loop $\ev{\tr P}$ as sample means of subset measurements \eqref{subsetmeas}.

\end{itemize}

\begin{figure}
\centerline{\includegraphics[height=0.15\textheight]{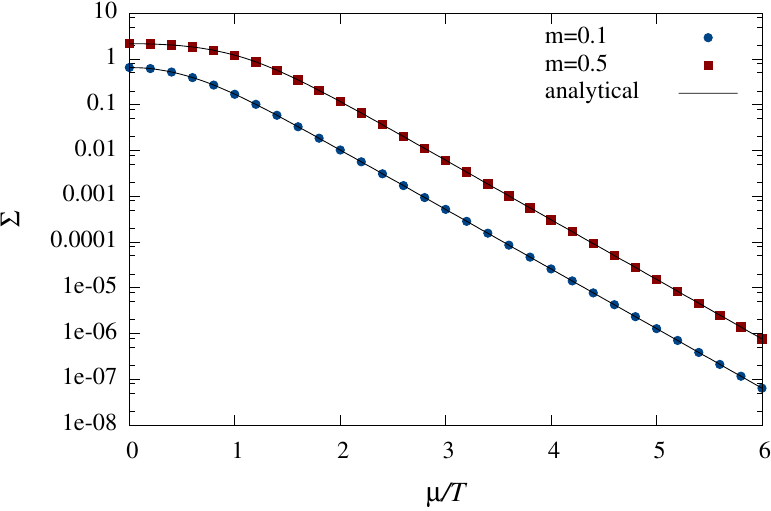} 
\includegraphics[height=0.15\textheight]{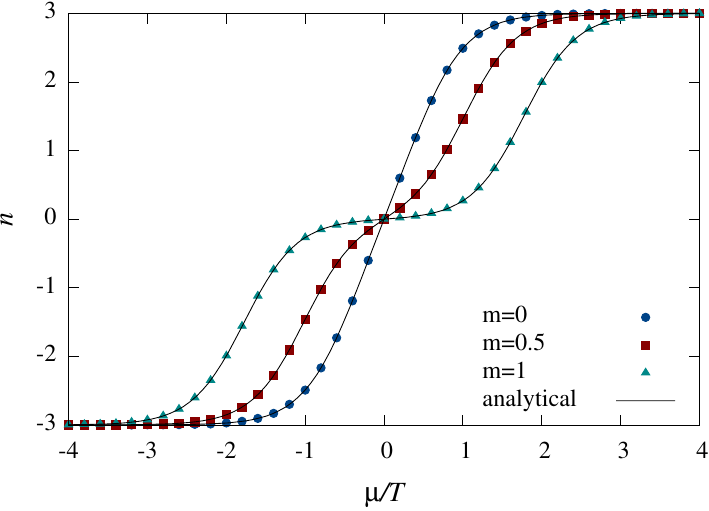}
\includegraphics[height=0.15\textheight]{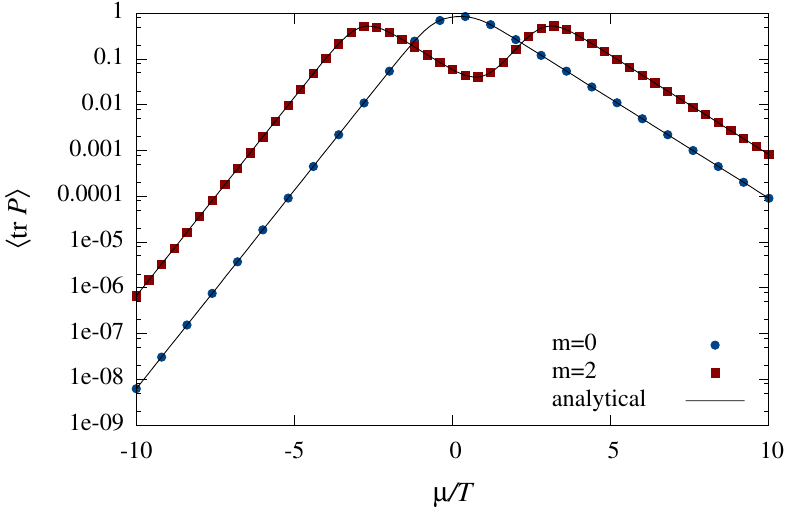}}
\caption{Chiral condensate $\Sigma$, quark number density $n$, and Polyakov loop $\ev{\tr P}$ as a function of $\mu$ for several values of the quark mass; computed from $N_\text{MC}=100,000$ subsets. }
\label{Fig:numres}
\end{figure}

In Fig.~\ref{Fig:numres} we show the results for the chiral condensate, quark number density, and Polyakov loop. The numerical results agree with the analytical predictions over several orders of magnitude. For the Polyakov loop we observe the $\mu \leftrightarrow -\mu$ asymmetry (or $\ev{\tr P} \leftrightarrow \ev{\tr P^\dagger}$ asymmetry), which is clearly illustrated by the different exponential decays for large positive and negative $\mu$.

\section{Larger $\mathbf{\emph{N}_\emph{f}}$ and extended subsets}

For arbitrary $N_f$ the subset weights are real by construction, but there is no general argument for their positivity. As a matter of fact the subset weights are only strictly positive for all $\mu$ and $P$ for $N_f<5.11$. For $N_f>5.11$ regions in $P$ and $\mu$ develop where the weights are negative. 

\begin{figure}
\centerline{\includegraphics[height=0.15\textheight]{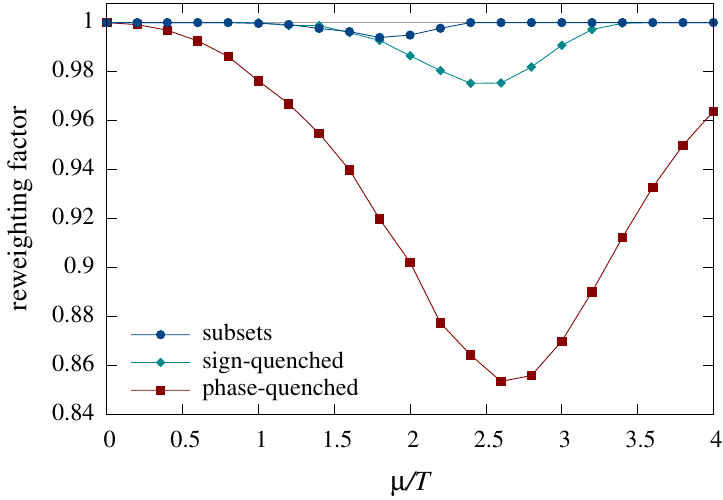}}
\caption{Reweighting factors for $N_f=12$ and $m=0$ in the subset formulation and in the phase-quenched and sign-quenched reweighting schemes in the link formulation.}
\label{Fig:rewfac}
\end{figure}

When the subset weights have a fluctuating sign, importance sampling can no longer be performed. 
Instead, one can use the subsets as an auxiliary system in reweighting. In this case the subsets are generated according to $|\sigma|$ and the sign of the weights is absorbed in the observable,
\begin{align}
    \ev O=\frac{\ev{\sign\sigma \times \ev{O}_{\Omega}}_{|\sigma|}}{\ev{\sign\sigma}_{|\sigma|}}\,.
 \label{rew}
\end{align}
In Fig.~\ref{Fig:rewfac} we compare the reweighting factors in the subset formulation with those in the phase-quenched and sign-quenched reweighting schemes in the original link formulation for $N_f=12$. Clearly, the sign problem is much milder in the subset formulation.

\begin{figure}
\centerline{
\includegraphics[width=0.3\textwidth]{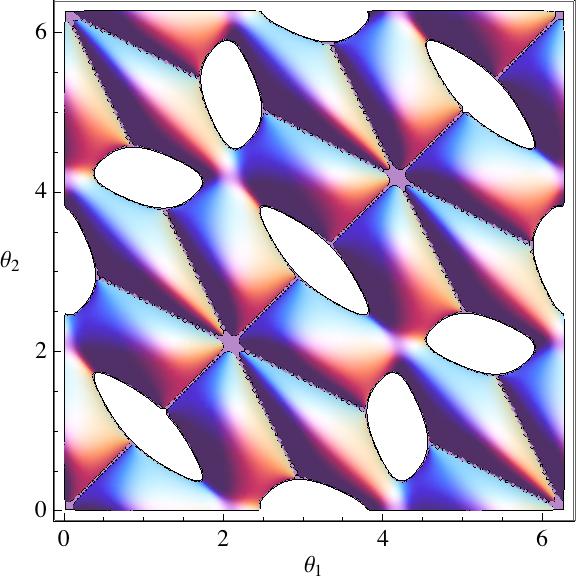}
\hspace{10mm}
\includegraphics[width=0.3\textwidth]{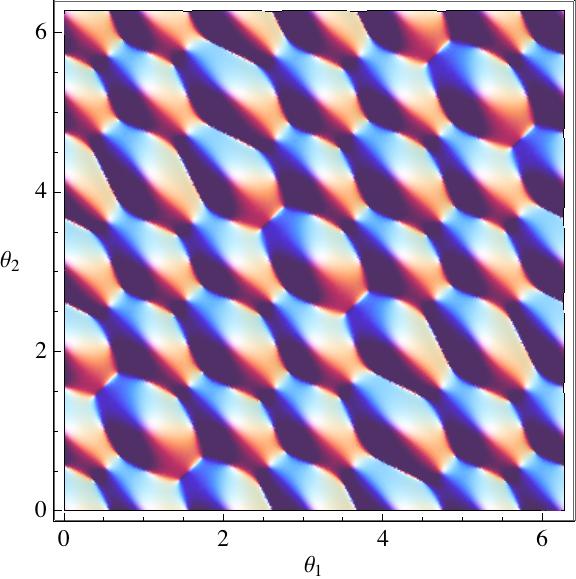}}
\caption{ Plot of $\log [J(P)\,\sigma(\Omega_P)]$ for the $Z_3$ subsets (left) and
$\log [J(P)\,\sigma^\text{ext}_P]$ for the extended subsets with $\alpha=-\beta=\pi/3$ (right) for $N_f=24$ ($m=0$) and $\mu/T=2.6$ with diagonal $P = \diag(e^{i\theta_1},e^{i\theta_2},e^{-i\theta_1-i\theta_2})$.}
\label{wext}
\end{figure}

Although we could always fall back onto the reweighting method, we now present an extension of the $Z_3$ subsets, which also have positive subset weights for larger $N_f$. We first analyze the weights of the $Z_3$ subsets. In Fig.~\ref{wext} (left) we show the product of the subset weights times the Haar measure for $N_f=24$ with $\mu/T=2.6$ for diagonal Polyakov loops characterized by their two independent eigenvalues $\theta_1$ and $\theta_2$. The data are plotted on a logarithmic scale such that the holes in the surface correspond to negative weights.
The mosaic of six replicated regions in the figure reveals the  permutation symmetries of $\theta_1$, $\theta_2$, and $\theta_3$.

We now construct extended subsets beyond $Z_3$ to solve the sign problem for $N_f \geq 6$.
First consider a constant diagonal \SU(3) matrix $G=\diag(e^{i\alpha},e^{i\beta},e^{-i\alpha-i\beta})$. 
For any link $P$, which can be diagonalized as $P = U \diag(e^{i\theta_1},e^{i\theta_2},e^{-i\theta_1-i\theta_2})\,U^\dagger$, we define a \textit{rotated} link
\begin{align}
R(P,G) = U \diag(e^{i\theta_1'},e^{i\theta_2'},e^{-i\theta_1'-i\theta_2'})\,U^\dagger \; \in\; \SU(3)
\end{align}
by rotating the eigenvalue matrix of $P$ by $G$, such that $\theta_{1}'=\theta_{1}+\alpha$ and $\theta_{2}'=\theta_{2}+\beta$.

To preserve the symmetry under the eigenvalue permutations we create 6 rotated links $P^{(i)}=R(P,\pi_i(G))$, $i=1,\ldots,6$, using all permutations $\{\pi_1,\ldots,\pi_6\}$ of the eigenvalues of $G$.
The extended subset for a link $P$ is the union $\Omega^\text{ext}_P = \bigcup_{i=0}^6\,\Omega_{P^{(i)}}$
of the $Z_3$ subsets \eqref{Z3subset} for $P^{(0)}=P$ and $P^{(1)},\ldots,P^{(6)}$, where the extended subset weight is given by
\begin{align}
  \sigma^\text{ext}_{P} = \frac17 \sum_{i=0}^6 \frac{J(P^{(i)})}{J(P)}
  \, \sigma(\Omega_{P^{(i)}}) 
  \label{sigmaext}
\end{align}
with $\sigma(\Omega_{P^{(i)}})$ the $Z_3$ subset weights \eqref{sigma} and 
$J$ the Jacobian of the reduced Haar measure \cite{Bloch:2013ara}.

From the location of the holes in Fig.~\ref{wext} (left) we can make a guesstimate for the shifts $\alpha$ and $\beta$ in $G$, in order to make the weights \eqref{sigmaext} positive. The weights corresponding to a specific choice of parameters are plotted in Fig.~\ref{wext} (right). Clearly the holes have disappeared showing that the extended subsets solve the sign problem for larger $N_f$ with a suitable $G$.

\section{Summary}

In this talk we presented a subset method to eliminate the sign problem in simulations of \qone at nonzero
chemical potential. For $N_f \leq 5$ we gathered the \SU(3) links and their complex conjugate into $Z_3$ subsets and found that the sum of fermion determinants is real and positive. For $N_f\geq 6$ the $Z_3$ subset weights can become negative, and we subsequently constructed extended subsets using additional SU(3) rotations which again yield positive weights. We demonstrated that the positivity of the subset weights allows for Monte Carlo simulations of \qone by subset sampling.

\section{Outlook: Subsets beyond $\mathbf{\emph{d}=1}$}

One can now wonder whether the subset method introduced for \qone can also be of use in higher dimensions $d$, where the sign problem is much more severe. A naive port to $d>1$ could consist of making a direct product of $Z_3$ subsets for (a subset of) all the temporal links on the lattice. For $N$ lattice sites the computational cost of such an algorithm would grow exponentially in the volume as $3^N$, while there is no a priori reason to believe that such subsets would actually alleviate the sign problem. 

In Table \ref{Tab:rewfac} we present some preliminary results obtained with direct product subsets for QCD in two dimensions with staggered quarks. We compare the average reweighting factors for (a) phase-quenched and (b) sign-quenched reweighting in the link formulation, with the subset reweighting factors for (c) a single collective $Z_3$ rotation of all temporal links on one time slice, (d) a direct product of $Z_3$ subsets for the temporal links of all spatial sites on one time slice, and (e) a direct product of $Z_3$ subsets for the temporal links on all lattice sites. Data were collected for $2\times N_t$ grids with $N_t=2,4,6,8$ and for a $4\times4$ and $6\times6$ grid, all for $N_f=1$ in the strong-coupling limit. As can be seen from the phase-quenched reweighting factor (a) the sign problem steadily grows as $N_t$ is increased. Whereas a collective $Z_3$ rotation (c) does not bring much improvement in the two-dimensional case, the direct product of $Z_3$ subsets on a single time slice (d) substantially improves on the sign problem. However, the truly surprising observation is that a direct product of $Z_3$ subsets for \textit{all} lattice sites (e) yields subset weights that are real and positive in all cases considered.

We also verified the effect of the gauge action on the reweighting factors for the $2\times6$ lattice by switching on $\beta$ and leaving the strong-coupling regime. The subset weights have to be modified to take into account the different values of the gauge action for the different subset elements, and the sign problem slowly reappears even for the full product subsets. Nevertheless, for $\beta=1,2,3,4,5$ the reweighting factor is 1.0, 1.0, 0.984(7), 0.964(13), and 0.972(17) respectively, so that the sign problem remains very mild, at least for these parameter values.

As a further test we also looked at the direct product subsets for small lattices in three- and four-dimensional QCD, even though the computational cost is huge even for small lattices. We observe with great interest that for $2^3$, $2^2\times4$, and $2^4$ lattices the product $Z_3$ subsets always give \textit{positive} weights (as was verified on samples of 200 configurations). 

Although the direct product of $Z_3$ subsets somehow points to a relation with the SU(3) singlet states in QCD, we do not yet understand why this induces positivity of the subset weights.
Moreover, the exponential growth of the subset size seems a serious objection to the usefulness of these subsets in numerical simulation. Nevertheless, the mere fact that the full direct product subsets have positive weights is surprising and worth exploring further. That this positivity was not to be expected a priori can also be argued from the fact that such absolute positivity is absent in the loop formulation of Ref.~\cite{Karsch:1988zx}, even though the sign problem is very much reduced after performing the integration over the gauge links \cite{Fromm:2009xw}.

Work is in progress to derive a formal positivity proof for the weights of the product subsets and construct a method allowing for the computation of the subset weights at non-exponential cost.

\begin{table}
\begin{center}
\small
\begin{tabular}{|c|c|l|l|l|l||l||l|}
\hline
& grid & 
\multicolumn{1}{c|}{$2\times2$} &
\multicolumn{1}{c|}{$2\times4$} &
\multicolumn{1}{c|}{$2\times6$} &
\multicolumn{1}{c||}{$2\times8$} & 
\multicolumn{1}{c||}{$4\times 4$} &
\multicolumn{1}{c|}{$6\times 6$}\\
\hline
a & phase-quenched & 0.8134(3) & 0.4361(4) & 0.233(2) & 0.130(2) &  0.295(1) & 0.0311(4)\\
b & sign-quenched & 0.9271(2) & 0.6150(5) & 0.355(3) & 0.203(2) & 0.442(2)&  0.043(4)\\
c & collective $Z_3$ & 0.9778(9) & 0.777(4) & 0.500(6) & 0.303(8) & 0.557(6)& 0.055(9)\\
d & $\otimes_{x} Z_3(x,0)$ & 1.0 & 0.9896(5) & 0.885(2) & 0.670(5) & 0.9973(2) &  0.726(4) \\
e & $\mathbf{\otimes_{xt} Z_3(x,t)}$ & \textbf{1.0}   & \textbf{1.0}  &  \textbf{1.0}$^*$ & \textbf{1.0}$^*$ & \textbf{1.0}$^{*}$ & \multicolumn{1}{c|}{N/A} \\
\hline
\end{tabular}
\caption{Reweighting factors for 2d-QCD for $N_f=1$ ($m=0$) for (a) phase-quenched and (b) sign-quenched reweighting in the link-formulation, and for (c) a single collective $Z_3$ rotation on all temporal links on one time slice, (d) a direct product of $Z_3$ subsets for the temporal links of all spatial sites on one time slice and (e) a direct product of $Z_3$ subsets for the temporal links on all lattice sites. The columns give the data for $2\times N_t$ grids with $N_t=2,4,6,8$ and for a $4\times4$ and $6\times6$ grid, all in the strong-coupling limit at $\mu=0.3$ with $\Nmc=100,000$ ($^*$ means $\Nmc=1,000$).\vspace{-2ex}}
\label{Tab:rewfac}
\end{center}
\end{table}

\enlargethispage{0.5\baselineskip}
\sloppypar
\bibliography{biblio}
\bibliographystyle{JBJHEP}

\end{document}